\documentstyle [aps,psfig,multicol]{revtex}

\begin{document}

\title{Coulomb drag in compressible quantum Hall states} 

\author{Iddo Ussishkin and Ady Stern}

\address{Department of Condensed Matter Physics, The Weizmann
  Institute of Science, Rehovot 76100, Israel}

\date{January 20, 1997}

\maketitle

\begin{abstract}
  We consider the Coulomb drag between two layers of two-dimensional
  electronic gases subject to a strong magnetic field. We first focus
  on the case in which the electronic density is such that the Landau
  level filling fraction $\nu$ in each layer is at, or close to,
  $\nu=1/2$. Discussing the coupling between the layers in purely
  electronic terms, we show that the unique dependence of the
  longitudinal conductivity on wave-vector, observed in surface
  acoustic waves experiments, leads to a very slow decay of density
  fluctuations. Consequently, it has a crucial effect on the Coulomb
  drag, as manifested in the transresistivity
  $\rho_{\scriptscriptstyle D}$.  We find that the transresistivity is
  {\em very large} compared to its typical values at zero magnetic
  field, and that its temperature dependence is unique --
  $\rho_{\scriptscriptstyle D}\propto T^{4/3}$.  For filling factors
  at or close to $1/4$ and $3/4$ the transresistivity has the same
  $T$-dependence, and is larger than at $\nu = 1/2$.  We calculate
  $\rho_{\scriptscriptstyle D}$ for the $\nu=3/2$ case and propose
  that it might shed light on the spin polarization of electrons at
  $\nu=3/2$. We compare our results to recent calculations of
  $\rho_{\scriptscriptstyle D}$ at $\nu=1/2$ where a composite fermion
  approach was used and a $T^{4/3}$-dependence was obtained. We
  conclude that what appears in the composite fermion language to be
  drag induced by Chern-Simons interaction is, physically, electronic
  Coulomb drag.
\end{abstract}

\begin{multicols}{2}

\section{Introduction and summary of results}

The advancement of fabrication and lithography techniques of
semi-conductors have made it possible to investigate bi-layer systems
made of two electronic two-dimensional layers at very close proximity
(several hundred Angstroms) in a very controlled manner. Among the
controllable parameters are the tunneling of electrons between the
layers, the density of electrons at each layer, and external
parameters such as the magnetic field and temperature. Moreover, it is
also possible to make separate electrical contacts to the two layers.
Experiments carried out on such systems have revealed a wealth of
interesting phenomena, and have also stimulated an active theoretical
study \cite{MD90,GM96,E96,Bon,H96}.

One of the measurements performed on systems of two electronic layers
at close proximity is that of the transresistivity. For that
measurement a current $I_1$ is driven in one layer, while the current
in the other layer, $I_2$, is kept zero. Due to the interactions
between electrons at different layers, momentum is transfered from the
current carrying layer to the second one. As a consequence of that
momentum transfer, one needs to apply a voltage on the second layer,
$V_2$, to keep $I_2$ zero. The ratio $- V_2 / I_1$ is defined as the
transresistivity, $\rho_{\scriptscriptstyle D}$. The transresistivity
was measured experimentally in \cite{Experiments} and studied
theoretically at \cite{Theories,ZH93,KO95}. Transresistivity in a
magnetic field was studied in \cite{TheoriesB}.  It was found that
although $\rho_{\scriptscriptstyle D}$ is a $dc$ measurement, it
reflects the response of the two layers to driving forces of finite
wave vector $\bf q$ and finite frequency $\omega$. In the limit of
weak coupling between the layers, the resistivity was found to be
mostly due to the Coulomb interaction between electrons at the two
layers, and to be given by the following expression, {\it valid both
  in the absence and presence of a magnetic field} \cite{ZH93},
\end{multicols}
\widetext

\noindent
\begin{picture}(3.375,0)
  \put(0,0){\line(1,0){3.375}}
  \put(3.375,0){\line(0,1){0.08}}
\end{picture}
\begin{equation}
  \label{donald}
  \rho_{\scriptscriptstyle D} = \frac{1}{2(2\pi)^2}
  \frac{h}{e^2}\frac{1}{Tn_1n_2}
  \int \frac{d{\bf q}}{(2\pi)^2}\int_0^\infty  
  \frac{\hbar\, d\omega}{\sinh^{2}{\frac{\hbar\omega}{2T}}} \,
  q^2\, |U_{\rm sc}({\bf q},\omega)|^2\, 
  {\rm Im} \Pi^{(1)}({\bf q},\omega){\rm Im} \Pi^{(2)}({\bf
    q},\omega) \, ,
\end{equation}
\hfill
\begin{picture}(3.375,0)
  \put(0,0){\line(1,0){3.375}}
  \put(0,0){\line(0,-1){0.08}}
\end{picture}

\begin{multicols}{2}
\noindent
where $U_{\rm sc}$ is the screened inter-layer Coulomb interaction,
$\Pi^{(i)}$ is the single-layer density-density response function of
the $i$'th layer, $n_i$ is the average density of electrons at layer
$i$ and $T$ is the temperature. Both $U_{\rm sc}$ and $\Pi$ are
defined precisely in the next section, where Eq.\ (\ref{donald}) is
studied in detail. There we also comment on the validity of Eq.\ 
(\ref{donald}) in the presence of a magnetic field.

In this paper we consider the transresistivity, in the limit of weak
inter-layer coupling, of two systems of two dimensional electron gases
(2DEG) in a strong magnetic field, where the Landau level filling
fractions of the two layers, $\nu_1$ and $\nu_2$, are smaller than one
and the electrons form a compressible state. The most prominent
example is that of $\nu_1$ and $\nu_2$ being equal to, or close to,
$1/2$.  In Section \ref{threehalf} we consider the $\nu_1=\nu_2=3/2$
case.  Our work is motivated by several experimental and theoretical
works that discovered a unique response of the $\nu=1/2$ state to
finite ${\bf q},\omega$ driving forces. A preliminary experimental
study of Coulomb drag in the regime of a partially filled Landau level
was recently carried out by Eisenstein {\em et al.} \cite{new}.

Most theoretical studies of the single layer $\nu=1/2$ state have used
the composite fermions approach, as introduced in
\cite{CFGeneral,HLR93}, to calculate electronic quantities, such as
the electronic linear response functions. In our study of the
transresistivity we use the composite fermion approach to obtain
information regarding the electronic single layer density-density
response function, {\it but discuss the coupling between the layers in
  purely electronic terms}. This approach makes the physical picture
underlying the transresistivity quite transparent.  In particular, it
allows us to point out that the unique features of the
transresistivity in the case we study are closely related to the
unique $q$-dependence of the electronic longitudinal conductivity, as
observed in surface acoustic waves measurements. Both phenomena
reflect the slow relaxation of density fluctuations in the $\nu=1/2$
state.

\begin{figure}
  \narrowtext
  \setlength{\unitlength}{2.8in}
  \vspace{0.8cm}
  \begin{picture}(1,0.618)
    \put(0,0){\psfig{figure=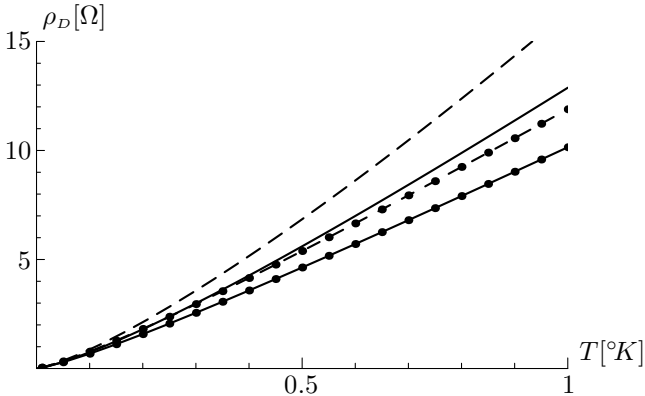,width=2.8in}}
    \put(1.02,0.02){\makebox(0,0)[lb]{$T [ ^\circ \! K ]$}}
    \put(0,0.648){\makebox(0,0)[bl]{
        $\rho_{\scriptscriptstyle D} [ \Omega ]$}}
    \put(-0.005, 0.206){\makebox(0,0)[r]{$5$}}
    \put(-0.005, 0.412){\makebox(0,0)[r]{$10$}}
    \put(-0.005, 0.618){\makebox(0,0)[r]{$15$}}
    \put(0.5, -0.01){\makebox(0,0)[t]{$0.5$}}
    \put(1, -0.01){\makebox(0,0)[t]{$1$}}
  \end{picture}
  \vspace{0.4cm}
  \caption{
    The transresistivity as a function of temperature for $\nu=1/2$,
    $n=1.4\times 10^{11}$cm$^{-2}$, and $d=300A$. The solid lines are
    calculations assuming $m^*=4m_b$, while the dashed lines assume
    $m^*=12m_b$. The corresponding values of $\alpha$ are $0.11$ and
    $0.037$. The dotted lines are numerical calculations, while the
    undotted ones are plots of Eq.\ (\protect{\ref{onehalf}}). }
  \label{plot1}
\end{figure}

As discussed in detail below, based on the use of Eq.\ (\ref{donald}),
with the $\Pi$ and $U_{\rm sc}$ appropriate for the case at hand, we
find,
\begin{enumerate}
\item For $\nu_1=\nu_2=1/2$, the leading order temperature dependence
  of the transresistivity is
  \begin{eqnarray}
    \rho_{\scriptscriptstyle D} & = &  0.825 \frac {h}{e^2}  
    \left( \frac {T}{T_0} \right)^{\frac {4}{3}}  \nonumber \\ 
    & - & 0.1775 \frac {h}{e^2}
    \frac {1 + 12 \alpha + 6 \alpha^2}{1 + \alpha} (4\pi n)^{1/2} d
    \left( \frac {T}{T_0} \right)^{ \frac {5}{3}}
    \nonumber \\ & + &
    {\cal O}(T^2\log T) \, ,
    \label{onehalf}
  \end{eqnarray}
  with
  \begin{equation}
    \label{4:T0}
    T_0 = \frac {4\pi e^2 n d}{{\tilde\phi}^2\epsilon}
    (1 + \alpha) \, .
  \end{equation}
  In Eq.\ (\ref{4:T0}) $\tilde \phi=2$ for $\nu=1/2$ and $\alpha^{-1}
  \equiv \frac{2\pi e^2d}{\epsilon} \frac{dn}{d\mu}$ where
  $\frac{dn}{d\mu}$ is the thermodynamical compressibility of the
  $\nu=1/2$ state and $\epsilon$ is the dielectric constant.
  Essentially, $\alpha$ is the ratio of the Thomas-Fermi screening
  length of the $\nu=1/2$ state to the separation between the layers,
  $d$.  For realistic numbers, $\alpha$ is small compared to one. In
  other words, the screening length is smaller than $d$.  The
  temperature scale $T_0$ is of the order of $190^\circ \! K$. A plot
  of the calculated $\rho_{\scriptscriptstyle D}$ as a function of
  temperature, at $\nu=1/2$ is given in Fig.\ \ref{plot1}.
  
  The effect of disorder is neglected in the derivation of Eq.\ 
  (\ref{onehalf}), and this neglect is justified as long as $\rho_{xx}
  \ll 2 \frac{h}{e^2} (T / T_0)^{1/3}$, where $\rho_{xx}$ is the
  single layer diagonal resistivity, and as long as $(T / T_0)^{1/3}
  \ll 1$.
  
  The transresistivity at $\nu=1/2$, as given by Eq.\ (\ref{onehalf}),
  is {\em much larger} than at $B=0$. The latter is given by
  \cite{KO95}
  \begin{equation}
    \rho_{\scriptscriptstyle D} (B=0) = 0.118\frac{h}{e^2}
    \left (\frac{T\epsilon}{e^2
        k_{\scriptscriptstyle F}^3 d^2}\right )^2 \, ,
    \label{bez}
  \end{equation} 
  where $k_{\scriptscriptstyle F}$ is the Fermi wavevector.
  Interestingly, in the limit of strong screening, both
  $\rho_{\scriptscriptstyle D}$ at $B=0$ and $\rho_{\scriptscriptstyle
    D}$ at $\nu=1/2$ are determined by a combination of geometrical
  factors (interlayer distance and inter-particle distance) and the
  strength of the Coulomb interaction. Both are independent of the
  mass, whether bare or effective.
  
  Eq.\ (\ref{onehalf}) holds also for the $\nu=1/4$ and $\nu=3/4$
  cases, with one modification, namely $\tilde\phi=4$. Thus, at low
  $T$, if the filling fraction is varied from $\nu=1/2$ to $\nu=1/4$
  or $\nu=3/4$, keeping the electron density fixed, the
  transresistivity is expected to increase by a factor smaller, but
  close to, $4^{4/3}$.
  
  While this paper was in preparation, two other related works were
  submitted for publication \cite{KM96,S96}. Both have considered the
  case $\nu_1=\nu_2=1/2$ and have found a temperature dependence of
  $\rho_{\scriptscriptstyle D}\propto T^{4/3}$. Our analysis agrees
  with this result.  Both works have used a composite fermions
  approach to analyze the transresistivity, and in one the unique
  temperature dependence was interpreted as reflecting the non-Fermi
  liquid properties of the composite fermions. As explained below, our
  understanding of this temperature dependence is different.
  
\item For $\nu_1=\nu_2=1/2$ we study the temperature dependence of
  $\rho_{\scriptscriptstyle D}$ for electron-electron interaction of
  the form $V(r)\propto \frac{1}{|r|^\eta}$.  Although this study is
  mostly of theoretical value, it was proven to be of interest in the
  subject of the composite fermion Fermi liquid theory
  \cite{HLR93,CFFLT}. We find the leading term to be,
  \begin{equation}
    \rho_{\scriptscriptstyle D}\propto \left\{ 
      \begin{array}{ll}
        T^{\frac{4}{2+\eta}} \ \ \ &{\rm for} \ \ 0<\eta<1 \\
        T^{4/3}                    &{\rm for} \ \ 1<\eta<2 
      \end{array}
    \right. \, .
    \label{range}
  \end{equation}
  
\item For the case of $\nu_1=\nu_2$ and both near $1/2$ we find the
  leading temperature dependence of $\rho_{\scriptscriptstyle D}$ to
  be $T^{4/3}$.  Based on a semiclassical approximation, we find that
  this statement holds as long as the composite fermion cyclotron
  radius in each layer, $R_c^*\propto |\nu-1/2|^{-1}$, is larger than
  the distance a composite fermion traverses without being scattered
  by impurities.  The effect of fermion-fermion scattering on
  $\rho_{\scriptscriptstyle D}$ is presently under study.
  
\item For two layers of densities $n_1={\bar n}+\frac{1}{2}\Delta n$
  and $n_2={\bar n}-\frac{1}{2}\Delta n$, with the average density
  corresponding to $\nu=1/2$, we find that for low temperature and
  small $\Delta n$,
  \begin{equation}
    \rho_{\scriptscriptstyle D}(\Delta n) =
    \rho_{\scriptscriptstyle D}(\Delta n=0) \left[
      1+\frac{7}{48}\left(\frac{\Delta
          n}{\bar n}\right )^2 \right] \, .
    \label{dn}
  \end{equation} 
  
\item Finally, we study the case of $\nu_1=\nu_2=3/2$.  The spin
  polarization of the single layer $\nu=3/2$ state is, at the moment,
  not well understood \cite{Stormer}. We calculate
  $\rho_{\scriptscriptstyle D}$ for the cases of complete spin
  polarization and zero spin polarization, and find that the
  transresistivity for a completely unpolarized $\nu=3/2$ state is
  larger by a factor close to $2^{2/3}$ compared to the
  transresistivity of a fully polarized $\nu=3/2$ state.
  
  Thus, if the picture emerging from Ref. \cite{Stormer} is correct,
  and the $\nu=3/2$ is not spin-polarized, but may become polarized by
  an application of a magnetic field $B_{||}$ {\it parallel} to the
  layer, a measurement of $\rho_{\scriptscriptstyle D}$ as a function
  of $B_{||}$ should be sensitive to that change of polarization.
  
  By passing, we also note that a change of polarization should be
  reflected also in surface acoustic waves measurements similar to the
  ones reported by Willet \cite{Wil94}.
\end{enumerate}

Before turning to describe the physical picture emerging from our
study, we comment on the approximations made. Eq.\ (\ref{donald}) is
derived under the assumption of weak inter-layer coupling. All
throughout this paper we neglect the finite thickness of the two
layers. Since our main findings result from small-$q$ behavior, we
expect the finite thickness not to affect any of the qualitative
features outlined above, but rather to modify prefactors only. Both of
these approximations are conventionally taken in Coulomb drag studies
\cite{Theories}.

Approximations special to the case at hand are taken when the single
layer $\Pi({\bf q},\omega)$ is calculated. As explained below, we
carry our calculations in the modified RPA (mRPA) approach to the
composite fermions Fermi liquid theory \cite{SH93}. We find, however,
that our main findings turn out to be independent of the composite
fermion effective mass $m^*$ and Landau parameter $f_1$ introduced
within mRPA, and thus can be expected to be insensitive to the details
of the approximation.

The structure of the paper is as follows: in Section \ref{physicalP}
we describe the physical picture behind our results. In Section
\ref{calculations} we give the detailed calculations. Section
\ref{summary} summarizes the paper. In the appendix we discuss the
relation of our approach to the composite fermion approach to the
problem, as employed in \cite{KM96} and \cite{S96}.

\section{The Physical Picture}
\label {physicalP}

In this section we describe the physical picture behind the results
summarized above. We try to point out clearly what features of our
analysis are independent of the magnetic field, and, in contrast, what
is unique to $\nu=1/2$ and $\nu$ close to $1/2$. To this end, we start
by reviewing the general theoretical considerations leading to Eq.\ 
(\ref{donald}) and the precise definition of each term in this
equation. After doing that, we focus on the $\nu=1/2$, and show that
the unique features we find in the transresistivity result from the
slow relaxation of density fluctuations in that state.

\subsection{General theoretical arguments regarding the transresistivity}

The physical picture behind Eq.\ (\ref{donald}) was discussed by
various authors \cite{Theories}. Here, we elaborate on some points
that are essential to our discussion.  Drag resistivity stems from
scattering events between electrons of different layers. This
scattering results from the screened inter-layer Coulomb interaction
$U_{\rm sc}$. Scattering events transfer momentum $\hbar{\bf q}$ and
energy $\hbar\omega$ between the layers. Phase space availability
effectively constrains the energy transfer to be smaller than the
temperature, as implied by the $\sinh^{-2}{\frac{\hbar\omega}{2T}}$
factor.  The function $\Pi({\bf q},\omega)$ is the single layer
density-density response function, irreducible with respect to the
Coulomb interaction. It is defined in the following manner: suppose
that one applies a weak scalar potential $V^{\rm ext}({\bf q},\omega)$
on a single layer (in the absence of a second one). In linear response
this potential leads to a particle density response $\rho({\bf
  q},\omega)$, which induces a Coulomb potential $V^{\rm
  ind}=\frac{2\pi e^2}{q}\rho({\bf q},\omega)$. The response function
$\Pi$ is the ratio of the induced density to {\it total} electric
potential $V^{\rm tot}\equiv V^{\rm ext}+V^{\rm ind}$,
\begin{equation}
  \label{pidef}
  \rho({\bf q},\omega)=-\Pi({\bf q},\omega)V^{\rm tot}({\bf q},\omega) \, .
\end{equation}
Since current conservation implies $\ \omega\rho={\bf q}\cdot {\bf J}$
and the electric field is related to the potential by a gradient
operator, the longitudinal conductivity $\sigma({\bf q},\omega)$,
relating charge current to the electric field, is related to $\Pi({\bf
  q},\omega)$ through
\begin{equation}
  \sigma({\bf
    q},\omega)=- i e^2 \frac{\omega}{q^2}\Pi({\bf q},\omega) \, .
  \label{pisigma}
\end{equation}

The response function $\Pi({\bf q},\omega)$ also determines the
screening of the Coulomb potential, both inter-layer and intra-layer.
We denote the bare intra-layer Coulomb interaction by $V_b(q)$ and the
bare inter-layer one by $U_b(q)$. It is convenient to describe the
interaction by $2\times 2$ matrices, where the two entries correspond
to the two layers. The interlayer screened Coulomb interaction is the
off-diagonal element of the matrix ${\hat V}_{\rm sc}$.  The latter
satisfies the equation
\begin{equation}
  \label{screened}
  {\hat V}_{\rm sc}={\hat V}_{\rm bare}
  \frac{1}{1+{\hat \Pi}{\hat V_{\rm bare}}} \, ,
\end{equation}
with 
\begin{equation}
  \label{matrixdef}
  {\hat V}_{\rm bare}\equiv\left(
    \begin{array}{cc}
      V_b & U_b\\
      U_b & V_b
    \end{array}
  \right )  \; \; \; {\rm and} \; \; \; 
  {\hat \Pi}\equiv \left (
    \begin{array}{cc}
      \Pi_1 & 0\\
      0 & \Pi_2
    \end{array}
  \right) \, .
\end{equation}
Note the notation: ${\hat \Pi}, {\hat V}$ are matrices, $\Pi_{1(2)}$,
$V_b$ and $U_b$ are scalar functions.

We now limit ourselves to the case of two identical layers, which is
our main focus in this paper. For that case $\Pi_1=\Pi_2$, such that
the subscript may be omitted. Then, the screened inter-layer potential
can be written as
\begin{equation}
  U_{\rm sc} = \frac{1}{2} 
  \frac{V_b+U_b}{1+\Pi({\bf q},\omega)(V_b+U_b)}
  -\frac{1}{2}\frac{V_b-U_b}{1+\Pi({\bf q},\omega)(V_b-U_b)} \, .
  \label{interscg}
\end{equation}
Eq.\ (\ref{interscg}) can be understood in the following way.  The
screened potential is the potential induced in one layer by a test
charge of magnitude one in the other layer, together with the
screening cloud it creates. One can view a test charge in one layer as
a sum of two charge distributions---a symmetric distribution, with the
same test charge in both layers, and an antisymmetric distribution,
with opposite test charges in the two layers.  The bare Coulomb
potential induced by a symmetric distribution is $V_b+U_b$, while the
bare potential induced by an anti-symmetric distribution is $V_b-U_b$.
The two terms in (\ref{interscg}) are the screened Coulomb potentials
induced by the symmetric and anti-symmetric distributions.  For the
case we consider here, of two coupled two-dimensional systems,
$V_b=\frac{2\pi e^2}{q}$ and $U_b=\frac{2\pi e^2}{q} e^{-qd}$, such
that, for $qd\ll 1$, $V_b+U_b\approx 2V_b$ and $V_b-U_b\approx V_bqd$.
A symmetric distribution induces a screened Coulomb potential, while
an anti-symmetric one induces a screened Coulomb potential of {\it
  electric dipoles}, whose dipole moment is $ed$.

We now focus our attention on the product ${\rm Im} \Pi({\bf
  q},\omega)|U_{\rm sc}({\bf q},\omega)|$, a product that appears in
Eq.\ (\ref{donald}). Using Eq.\ (\ref{interscg}), and suppressing, for
brevity, the ${\bf q},\omega$ arguments of $V_b,U_b$ and $\Pi$, this
product may be written as
\begin{eqnarray}
  \lefteqn {{\rm Im} \Pi({\bf q},\omega) 
    |U_{\rm sc}({\bf q},\omega)| = } 
  \nonumber \\ & &
  -{\rm Im} \Pi^{-1}\ \left|\frac {U_b}{(\Pi^{-1}+V_b+U_b)
      (\Pi^{-1}+V_b-U_b)} \right| \, .
  \label{impiv}
\end{eqnarray}
By Eq.\ (\ref{donald}), the transresistivity will be determined by
${\bf q},\omega$ for which the product ${\rm Im} \Pi({\bf
  q},\omega)|U_{\rm sc}({\bf q},\omega)|$ is large. Thus, it is
interesting to examine its poles. As we see below, ${\rm Im}
\Pi^{-1}({\bf q},\omega)$ does not have any poles, and therefore the
only poles are the solutions to the equations,
\begin{equation}
  i\omega-\frac{q^2}{e^2}\sigma({\bf q},\omega)(V_b(q)+U_b(q)) = 0 
  \label{sympole}
\end{equation}
\begin{equation}
  i\omega-\frac{q^2}{e^2}\sigma({\bf q},\omega)(V_b(q)-U_b(q)) = 0 
  \label{asympole} \, .
\end{equation}
To illuminate the physical significance of the poles, we expressed
$\Pi$ in terms of the conductivity $\sigma$.  The solutions to
(\ref{sympole}) and (\ref{asympole}) are the dispersion relations for
charge density modes: Suppose that at time $t=0$ one sets up a
symmetric charge modulation, $\rho_0({\bf q})$ in each of the two
layers, and observes its time evolution.  This charge modulation
induces, in each of the layers, an electric field given by ${\bf\cal
  E}=-i{\bf q}\rho({\bf q},t)(V_b(q)+U_b(q))$.  The electric field
generates, in turn, , a current ${\bf j}=\sigma {\bf\cal E}$ in each
of the layers.  Since $\frac{\partial\rho}{\partial t} =-i{\bf
  q}\cdot{\bf j}$, the current leads to a decay of the charge
modulation. The dispersion relation corresponding to that decay is the
one given in Eq.\ (\ref{sympole}).  The second pole results from an
anti-symmetric charge distribution, i.e., a distribution in which at
$t=0$ the charge modulation in the two layers is $\pm\rho({\bf q})$.
In that case, the electric field induced in the two layers is
$\pm{\bf\cal E}=-i{\bf q}\rho({\bf q},t)(V_b(q)-U_b(q))$. Again, this
electric field leads to a decay of the charge modulation, following
the dispersion relation given in Eq.\ (\ref{asympole}).

In both cases, when $\sigma$ is real, the density modes are damped,
and density modulations decay in time. When $\sigma$ is purely
imaginary, the solutions to Eqs. (\ref{sympole}) and (\ref{asympole})
are the plasma modes of the double layer system.

\subsection{What is unique in the transresistivity of the $\nu=1/2$ case?}

As pointed out by Zheng and MacDonald \cite{ZH93}, Eq.\ (\ref{donald})
is valid both in the absence and in the presence of a magnetic field.
A simple way of understanding that point is by following the
derivation of (\ref{donald}), as given by Zheng and MacDonald. Upon
doing that, one finds that the derivation is unchanged in the presence
of a magnetic field.  Thus, the study of drag resistivity for the
system at hand is reduced to a study of the density-density response
function $\Pi$ for electrons at $\nu=1/2$.  And it is there where the
behavior of electrons at $\nu=1/2$ differs strongly from that of
electrons at $B=0$.
 
The calculation of $\Pi({\bf q},\omega)$ is carried out, following a
similar calculation by Halperin, Lee and Read (HLR)\cite{HLR93}, in
the next section, and is summarized by Eq.\ (\ref{pizz}). For the
description of the physical picture it is enough to quote the result
for $q\ll k_{\scriptscriptstyle F}$ and $\omega\ll
qv_{\scriptscriptstyle F}$ (with $v_{\scriptscriptstyle F}$ being the
Fermi velocity):
\begin{equation}
  \Pi({\bf q},\omega)\approx \frac{q^3}{q^3\left( 
      \frac{dn}{d\mu}\right)^{-1}-2\pi i\hbar{\tilde\phi}^2\omega
    k_{\scriptscriptstyle F}} \, ,
  \label{Pieq}
\end{equation}
where $\Pi(q\rightarrow 0,\omega=0)\ \equiv\ \frac{dn}{d\mu}$ is, by
definition, the thermodynamic electronic compressibility of the
$\nu=1/2$ state, and $k_{\scriptscriptstyle F}\equiv\sqrt{4\pi n}$.

This form of $\Pi$ is unique, and deserves a few comments: firstly, it
is similar to the form of $\Pi$ for a diffusive system at $B=0$,
namely, $\frac{dn}{d\mu}\frac{Dq^2} {-i\omega + Dq^2}$, but with the
effective "diffusion constant" being linear in $q$. Eq.\ (\ref{Pieq})
describes, then, a very slow diffusion, where the diffusion constant
vanishes for $q\rightarrow 0$. This slow diffusion affects the
dispersion relation of the two density relaxation poles of the
screened Coulomb potential, defined in Eqs.  (\ref{sympole}) and
(\ref{asympole}). Those become $\omega\propto q^3[V_b(q)+U_b(q)]$ and
$\omega\propto q^3[V_b(q)-U_b(q)]$, and describe a slow relaxation of
density fluctuations, that eventually leads to a large drag
resistivity.  Secondly, as revealed by surface acoustic waves
experiments, explained by the composite fermion theory, and manifested
in Eq.\ (\ref{Pieq}), the $q$-dependence of the longitudinal
conductivity of electrons at $\nu=1/2$ is very different from that of
electrons at zero magnetic field. At $B=0$ the electronic conductivity
is usually at its maximum at $q=0$, and decreases with increasing $q$.
At $\nu=1/2$, in the absence of disorder, the longitudinal
conductivity {\it vanishes} at $q=0$, and increases linearly with
increasing $q$, as long as $q$ is not too large. Again, this small
longitudinal conductivity leads to a slow decay of density
fluctuations, and thus to a large transresistivity. Thirdly, Eq.\ 
(\ref{Pieq}) describes a strong suppression of ${\rm
  Im}\Pi^{-1}\propto 1/q^3$ for large values of $q$.  Consequently,
the $q$ integral in (\ref{donald}) is {\it not} dominated by the upper
cut-off $q\sim d^{-1}$, as is the case at $B=0$, but rather by the
solution $q_0(\omega)$ of Eq.\ (\ref{asympole}). Since the $\omega$
integral is dominated by $\omega\approx T$, the most important
contribution to the transresistivity comes from the region
\begin{equation}
  q\approx q_0(T)\approx k_{\scriptscriptstyle F}
  \left(\frac{T}{T_0}\right)^{1/3} \, .
  \label{qnot}
\end{equation}

Substituting Eq.\ (\ref{Pieq}) into Eqs. (\ref{impiv}) and
(\ref{donald}) we get the leading order of Eq.\ (\ref{onehalf}). The
detailed calculation, including the next two terms of Eq.\ 
(\ref{onehalf}), is given in the next section.

Eq.\ (\ref{qnot}) puts a limit on the validity of our results. Our use
of Eq.\ (\ref{Pieq}) is valid only for $q_0(T) \gg l_{el}^{-1}$, where
$l_{el}$ is the single layer composite fermion elastic mean free path
\cite{HLR93}. Since the electronic $\rho_{xx} = 2
\frac{h}{e^2}\frac{1}{k_{\scriptscriptstyle F} l_{el}}$, this
condition amounts to
\begin{equation}
  \rho_{xx} \ll 2 \frac{h}{e^2}
  \left (\frac{T}{T_0}\right )^{1/3} \, .
  \label{validity}
\end{equation}

With the above discussion in mind, the origin of the dependence of
$\rho_{\scriptscriptstyle D}$ on the range of the Coulomb interaction
[Eq.\ (\ref{range})] can be understood.  Interactions decaying slower
than $1/r$ lead to a faster relaxation of long wavelength density
fluctuations, and thus to a less singular temperature dependence of
$\rho_{\scriptscriptstyle D}$ (although still more singular than the
$T^2$-dependence of non-interacting electrons).  The case of short
range interactions is discussed below.

We now comment on the temperature dependence of the transresistivity
when the electron density in the two layers is varied such that the
filling fraction is slightly away from $\nu=1/2$.  The
transresistivity is again determined by Eq.\ (\ref{donald}), with the
crucial ingredient being the density-density response function $\Pi$.
For the analysis of $\Pi({\bf q},\omega)$ we resort to the composite
fermion approach.

For $\nu$ slightly away from $1/2$, a new length , the composite
fermion cyclotron radius, $R_c^*$, and a new frequency scale, the
composite fermion cyclotron frequency, $\omega_c^*$ are introduced. As
long as $R_c^*$ is larger than the transport mean free path $l_{tr}$,
introduced by disorder, the composite fermion dynamics is, to a large
extent, unaffected by the effective magnetic field they are subject
to. The qualitative ${\bf q},\omega$ dependence is then similar to
that of the $\nu=1/2$ case, and we expect $\rho_{\scriptscriptstyle
  D}\propto T^{4/3}$.  Moreover, even when $R_c^*$ is smaller than the
disorder-induced mean free path, scattering events between composite
fermions might suppress the sensitivity of the composite fermions to
the effective magnetic field they are subject to. We thus expect that
the $T^{4/3}$ temperature dependence may even extend to the region
$R_c^*<l_{tr}$.

\section{Calculations}
\label{calculations}

\subsection{Transresistivity for $\nu_1=\nu_2=1/2$}

The response function $\Pi$ for a single layer at $\nu=1/2$ was
calculated by HLR within the random phase approximation (RPA).
Improvements to that approximations were developed in Refs.
\cite{SH93,SSH96}. We now review the RPA calculation and its
improvements.

The calculation of the electronic density-density response function
necessitates a notation for single layer response functions. The
single layer response functions are $2\times 2$ matrices, whose
entries correspond to density and transverse current.  We urge the
reader to distinguish single layer response functions from the
$2\times 2$ matrix $\hat\Pi$ (defined in Eq.\ \ref{matrixdef}). In the
latter the two entries correspond to densities in the two layers. To
help this distinction, we denote single layer matrices by tildes,
while double layer matrices are denoted by hats.

The single layer electronic density-density response function is the
density-density element of the electronic single layer response
function ${\tilde\Pi}^e$.  The following discussion, concluded by Eq.\ 
(\ref{pizz}), is devoted to a calculation of that element.
Consequently, the two matrices we are about to define now, $\tilde C$
and $\tilde\Pi^{\mbox {\tiny CF}}$, are {\it single layer} matrices,
with entries corresponding to density and transverse current.

Defining the Chern-Simons interaction matrix,
\begin{equation}
  {\tilde C}=\left(
    \begin{array}{cc}
      0 & \frac{2\pi i\hbar{\tilde\phi}}{q} \\
      -\frac{2\pi i\hbar{\tilde\phi}}{q}  & 0
    \end{array} 
  \right) \, ,
  \label{cs}
\end{equation}
we write 
\begin{equation}
  ({\tilde\Pi}^{e})^{-1} = 
  {\tilde C}+({\tilde\Pi}^{\mbox {\tiny CF}})^{-1}
  \label{picf}
\end{equation}
where ${\tilde\Pi}_{\mbox {\tiny CF}}$ is the composite fermion
density-density response function, describing the response of the
composite fermions to the total scalar and vector potentials they are
subject to, including external, Coulomb and Chern-Simons
contributions.  Diagrammatically, ${\tilde\Pi}^{\mbox {\tiny CF}}$ is
the sum of all diagrams irreducible with respect to a single
Chern-Simons or Coulomb interaction line.

Within RPA, the composite fermion response function
${\tilde\Pi}^{\mbox {\tiny CF}}$ is that of non-interacting fermions
at zero magnetic field. It is therefore a diagonal matrix. To lowest
order in $q\over k_{\scriptscriptstyle F}$ and
$\omega/(qv_{\scriptscriptstyle F})$ (with $v_{\scriptscriptstyle F}$
being the Fermi velocity) its components are,
\begin{eqnarray}
  \nonumber
  \Pi^{\mbox {\tiny CF}}_{00} & \approx & \frac{m}{2\pi\hbar^2} 
  \\
  \Pi^{\mbox {\tiny CF}}_{11} & \approx &
  -\frac{q^2}{24\pi m} + i 
  \frac{\omega k_{\scriptscriptstyle F}}{2\pi\hbar q} \, ,
\label{picfrpaa}
\end{eqnarray}
where $m$ is the bare mass. The finite value of $\Pi_{00}^{\mbox
  {\tiny CF}}$ in that limit reflects the compressibility of the
system, while the limit of $\Pi_{11}^{\mbox {\tiny CF}}$ reflects
Landau diamagnetism (the real part) and Landau damping (the imaginary
part).

Improvements to RPA include the modified RPA (mRPA), in which the bare
mass is replaced by a renormalized mass $m^*$, and a Landau
interaction parameter $f_1$ is introduced \cite{SH93}, and the
magnetized modified RPA (mmRPA), in which an orbital magnetization is
attached to each fermion \cite{SSH96}.

Within mRPA, Eq.\ (\ref{picfrpaa}) is replaced by
\begin{eqnarray}
  \Pi^{\mbox {\tiny CF}}_{00} & \approx & \frac{m^*}{2\pi\hbar^2} 
  \nonumber \\
  \Pi^{\mbox {\tiny CF}}_{11} & \approx &
  -\frac{q^2}{24\pi m^*}+i
  \frac{\omega k_{\scriptscriptstyle F}}{2\pi\hbar q} \, .
  \label{picfrpab}
\end{eqnarray}
The attachment of magnetization does not affect the component
$\Pi_{00}^e$ of the matrix $\tilde\Pi^e$, which is the only one
relevant for Coulomb drag \cite{SSH96}.  Thus, we use Eqs.
(\ref{picf}) and (\ref{picfrpab}) to calculate the electronic
density-density response function $\Pi=\Pi^{e}_{00}$ obtaining,
\begin{eqnarray}
  \Pi({\bf q},\omega) & = &
  \frac{\Pi^{\mbox {\tiny CF}}_{00}}{1-\Pi^{\mbox {\tiny CF}}_{00} 
    \Pi^{\mbox {\tiny CF}}_{11} 
    \left(\frac{2\pi\hbar \tilde\phi}{q}\right)^2}
  \nonumber \\
  & = & \frac{\frac{m^*}{2\pi\hbar^2}}
  {1+\left(\frac{2\pi\hbar\tilde\phi}{q^2}\right)^2
    \frac{m^*}{2\pi\hbar^2}
    \left(\frac{q^2}{24\pi m^*} -
      \frac{i\omega k_{\scriptscriptstyle F}}{2\pi\hbar q}\right)}
  \label{pizz}
\end{eqnarray}
which is Eq.\ (\ref{Pieq}).  Note that the thermodynamic
compressibility is $\frac{dn}{d\mu}=\frac{m^*}{2\pi\hbar^2}\left
  (1+\frac{\tilde\phi^2}{12}\right)^{-1}$.  Substituting Eq.\ 
(\ref{pizz}) in Eqs. (\ref{impiv}) and (\ref{donald}) we find,
\begin{eqnarray}
  \rho_{\scriptscriptstyle D}  & = &  \frac{\Gamma
    \left( \frac {7}{3} \right)
    \zeta \left( \frac {4}{3} \right)}{3\sqrt{3}} 
  \frac {h}{e^2}  
  \left( \frac {T}{T_0} \right)^{\frac {4}{3}}
  \nonumber \\
  & - & \frac{\Gamma \left( \frac {8}{3} \right)
    \zeta \left( \frac {5}{3} \right)}{18} \frac {h}{e^2}
  \frac {1 + 12 \alpha + 6 \alpha^2}{1 + \alpha} (4\pi n)^{1/2} d
  \left( \frac {T}{T_0} \right)^{ \frac {5}{3}}
  \nonumber \\ & + & {\cal O}(T^2\log T) \, .
  \label{onehalfd}
\end{eqnarray}
The origin of the $T^{5/3}$ term is in the second order expansion of
the $e^{-qd}$ factor in the bare inter-layer Coulomb interaction.

A numerical evaluation of Eq.\ (\ref{donald}) using (\ref{pizz})
yields the temperature dependence plotted in Fig. (\ref{plot1}). The
parameters used in the numerical calculation are $n=1.4\times
10^{11}$cm$^{-2}$, $d=300A$, and $\nu=1/2$. The numerical calculation
uses Eq.\ (\ref{pizz}) for $\Pi$, but uses the exact value of $U_b$,
rather than a small $qd$ expansion.

The transresistivity at $\nu=1/4$ is obtained directly from Eqs.
(\ref{onehalfd}) and (\ref{4:T0}). The transresistivity at $\nu=3/4$
is obtained by regarding the $\nu=3/4$ state as a superposition of a
$\nu=1/4$ state of holes and a $\nu=1$ state of electrons.

\subsection{Transresistivity for $\nu_1=\nu_2=1/2$, as a function of
  the range of interactions}
 
In the study of the Chern-Simons Fermi liquid theory formed by
composite fermions in the $\nu=1/2$ state it is useful to consider
electron-electron interactions of the form
\begin{equation}
  V(r)=\lambda |r|^{-\eta}
  \label{eta}
\end{equation}
with $0<\eta<2\ $ \cite{CFFLT}.  For $\eta<1$, the composite fermions
form a conventional Fermi liquid---the effective mass and
quasi-particle residues are finite. At $\eta=1$, Coulomb interaction,
the Fermi liquid is ``marginal'', i.e., both quantities are
logarithmically singular. And for $\eta>1$, the effective mass
strongly diverges and the quasi-particle residue vanishes at the Fermi
level.  In all cases, however, the compressibility of the $\nu=1/2$
state is not singular \cite{CFFLT}.

The small $q$ behavior of the potential (\ref{eta}) is $V({\bf
  q})\propto \frac{1}{q^{2-\eta}}$, and consequently the bare
potential induced by anti-symmetric density modulations is
$V_b-U_b\propto q^{\eta-1}d$. The form of $\Pi({\bf q},\omega)$ given
in Eqs. (\ref{Pieq}) and (\ref{pizz}) does not depend on the
interaction potential. Thus, in the calculation of the drag
resistivity one has to distinguish between two cases:
\begin{enumerate}
\item Long range interactions where $0<\eta<1$: In that case the range
  of the interaction determines the dispersion relation of the
  anti-symmetric density relaxation modes (\ref{asympole}) to be
  $\omega\propto q^{2+\eta}$. The leading temperature dependence of
  the transresistivity is then $\rho_{\scriptscriptstyle D}\propto
  T^{\frac{4}{2+\eta}}$.  Although the composite fermions form a Fermi
  liquid, $\rho_{\scriptscriptstyle D}$ does not exhibit the $T^2$
  temperature dependence characteristic of non-interacting electrons
  at zero magnetic field. The difference stems from the slow decay of
  density modulations.
  
\item Short range interactions where $1<\eta<2$: In that case the
  dispersion relation of anti-symmetric density relaxation modes
  (\ref{asympole}) is not determined by the interaction, but rather by
  the compressibility---a density modulation induces a spatial
  dependence in the chemical potential, which creates a current, which
  relaxes the modulation of the density. The dispersion relation is
  $\omega\propto q^3$, independent of $\eta$. Consequently, the
  temperature dependence of $\rho_{\scriptscriptstyle D}$ is not
  determined by $\eta$, and is $\rho_{\scriptscriptstyle D}\propto
  T^{4/3}$.
\end{enumerate}

\subsection{Transresistivity  close to $\nu=1/2$}

When the two layers are at identical filling fractions close to
$\nu=1/2$, the density-density response functions are identical
$\Pi^{(1)}=\Pi^{(2)}$. Within the composite fermion approach, both can
be calculated by Eq.\ (\ref{picf}). The function $\tilde\Pi^{\mbox
  {\tiny CF}}$ is, in that case, the response function of composite
fermions at an effective magnetic field of $\Delta B=B-2\Phi_0 n$. We
limit ourselves here to relatively high temperatures, $T\gg
\hbar\omega_c^*$. In that regime a semiclassical random phase
approximation may be employed for a calculation of $\Pi({\bf
  q},\omega)$, as was done, e.g., by HLR and by Simon and
Halperin\cite{SH93}. The semiclassical composite fermion response
function $\tilde\Pi^{\mbox {\tiny CF}}$ is \cite{SH93,US97},
\begin{eqnarray}
  \Pi^{\mbox {\tiny CF}}_{00} & = & \frac{m}{2 \pi \hbar^2} \frac{1 - 
    \frac{\pi W}{\sin{\pi W}} 
    J_W(qR_c^*)J_{-W}(qR_c^*)}{1 - \frac{i}{\omega_c^*\tau}
    \frac{\pi}{\sin{\pi W}} J_W(qR_c^*)J_{-W}(qR_c^*)}
  \nonumber \\ 
  \Pi^{\mbox {\tiny CF}}_{01} & =
  & - i v_{\scriptscriptstyle F} \frac {m}{4 \pi \hbar^2}
  \frac{ \frac {\omega}{\omega_c^*} \frac {\pi}{\sin \pi W}
    \left[J_W(qR_c^*)J_{-W}(qR_c^*)\right]'}{1 - 
    \frac{i}{\omega_c^*\tau} \frac{\pi }{\sin{\pi W}}
    J_W(qR_c^*)J_{-W}(qR_c^*)}
 \nonumber \\
  \Pi^{\mbox {\tiny CF}}_{11} &  = &
  \frac {\omega^2}{q^2} \Pi^{\mbox {\tiny CF}}_{00}
  \nonumber \\ \lefteqn { -
  v_{\scriptscriptstyle F}^2 \frac{m}{2 \pi \hbar^2} 
  \frac{ \frac {\omega}{\omega_c^*}
    \frac {\pi}{\sin \pi W} J_{1 + W}(qR_c^*)J_{1 - W}(qR_c^*)}{1 -
      \frac{i}{\omega_c^*\tau} \frac{\pi }{\sin{\pi W}}
      J_W(qR_c^*)J_{-W}(qR_c^*)} \, ,}
\label{scpicf}
\end{eqnarray}
where $W\equiv\frac{\omega}{\omega_c^*}+\frac{i}{\omega_c\tau}$, $J_W$
is the Bessel function of order $W$, and a prime denotes
differentiation with respect to the argument.

In the limit of $\omega_c^*\tau_{tr}\ll 1$ (or, equivalently
$R_c>l_{tr}$) we find, using asymptotic expansion of the Bessel
functions, that the response functions (\ref{scpicf}) reduce, at the
relevant range of $q,\omega$ to the zero effective magnetic field
response functions (\ref{picfrpaa}). Thus, in that regime the leading
temperature dependence of the transresistivity is expected to be
approximately equal to that of the $\nu=1/2$ case.

There are several contributions to the composite fermion scattering
rate. The first is impurity scattering, which is independent of
energy. The second is mutual composite fermion scattering, which does
depend linearly on energy (at least at the $\nu=1/2$ case
\cite{HLR93,CFFLT}). The precise effect of the latter on the response
at finite $q$ is a delicate subject, which is presently under
investigation \cite{US97}.

We now turn to consider two layers of densities $n_1= \bar{n} +
\frac{1}{2} \Delta n$ and $n_2 = \bar{n} - \frac{1}{2} \Delta n$,
where the density $\bar {n}$ corresponds to a filling fraction
$\nu=1/2$.  Measurements of the dependence of the transresistivity on
a density bias between the two layers were done in the past as a way
of separating out Coulomb drag and phonon drag mechanisms. A
calculation of the dependence of $\rho_{\scriptscriptstyle D}$ on such
a density bias could therefore be useful for such measurements at
$\nu=1/2$.

Since the two layers are not identical, the screened inter-layer
Coulomb potential cannot be put in the form (\ref{interscg}), and is
rather given by
\begin{equation}
  U_{\rm sc}=\frac 
  {U_b}
  {[1+{\bar\Pi}(V_b+U_b)] 
    [{\bar\Pi}(V_b-U_b)]-\delta\Pi^2(V_b^2-U_b^2)} \, ,
  \label{interscgne}
\end{equation}
where $\Pi_{1(2)}={\bar\Pi}\pm\delta\Pi$. 

The density bias $\Delta n$ changes the density-density response
function $\Pi$ in two ways. The first, trivial, way, is through the
dependence of $\Pi$ on the electron density. The second way is by
changing the filling fraction. However, as shown above, at least
within the semiclassical approximation, if $R_c^*$ is larger than the
mean free path the composite fermion response functions at the
relevant range of $q,\omega$ are approximately insensitive to the
effective magnetic field $\Delta B$.  Thus, the main effect of $\Delta
n$ comes through the dependence of $\Pi$ on the density. Using
(\ref{donald}) and (\ref{interscgne}) with two different response
functions for the two layers, we find that for small $\Delta n$, and
to leading order in the temperature, the transresistivity is
\begin{equation}
  \rho_{\scriptscriptstyle D}(\Delta n) =
  \rho_{\scriptscriptstyle D}(\Delta n=0)\ 
  \left[1+\frac{7}{48}\left(\frac{\Delta
        n}{\bar n}\right )^2 \right] \, .
  \label{deltan}
\end{equation}

\subsection{Transresistivity for $\nu_1=\nu_2=3/2$}
\label{threehalf}

We now analyze the transresistivity for two layers of $\nu=3/2$ each.
Our motivation for studying this case stems from the ambiguous
experimental picture of spin polarization at the $\nu=3/2$ state
\cite{Stormer}.  While transport measurements at tilted magnetic
fields indicate only a partial spin polarization of the $\nu=3/2$
state, surface acoustic waves measurements do not seem to support such
a picture \cite{Stormer}. We therefore attempt to find out whether a
measurement of transresistivity can shed light on the spin
polarization of the $\nu=3/2$ state.

We regard the $\nu=3/2$ state as a superposition of a $\nu=2$ state of
electrons and a $\nu=1/2$ state of holes, a picture which is valid
when the electronic Landau level separation $\hbar\omega_c$ is much
larger than the temperature and the Coulomb energy scales, such that
Landau level mixing can be neglected. Under these conditions the
contribution of the electronic $\nu=2$ state to the single-layer
density-density response function $\Pi$ is exponentially small for the
relevant frequency range $\hbar\omega<T$, due to the energy gap
between the $n=0$ and $n=1$ electronic Landau levels. The
transresistivity is therefore almost solely due to the $\nu=1/2$ state
of the holes.

The spin polarization of the $\nu=1/2$ hole state may be anywhere
between zero and full polarization. So far we considered only fully
polarized states. When considering a partially polarized state, one
expects all relevant functions (${\hat\Pi}, {\hat V_{\rm bare}}, {\hat
  V_{\rm sc}}$ of Eqs. (\ref{screened}) and (\ref{matrixdef})) to
acquire spin indices. As a first step, then, a convenient choice of a
basis may be helpful. This choice emerges naturally when considering
the bare Coulomb interaction, even within one layer. We write the
intra-layer bare interaction as a $2\times 2$ matrix, where the two
entries are for densities of electrons with two spin states.

In the basis of eigenvectors of $\sigma_z$,
\begin{equation}
  {\hat V}_{\rm bare}=\frac{2\pi e^2}{q}\left (
    \begin{array}{cc}
      1 & 1\\
      1 & 1
    \end{array}\right )
  \label{barespin}
\end{equation}
since the Coulomb interaction does not depend on the spin direction.
The convenient basis is the basis of charge density
$\rho_\uparrow+\rho_{\downarrow}$ and spin density
$\rho_\uparrow-\rho_{\downarrow}$ (where $\rho_\uparrow$
($\rho_{\downarrow}$) denote the density of electrons whose spin is
parallel (anti-parallel) to the magnetic field). In that basis,
\begin{equation}
  {\hat V}_{\rm bare}=\frac{2\pi e^2}{q}\left (
    \begin{array}{cc}
      1 & 0\\
      0 & 0
    \end{array}\right) \, .
  \label{barespinb}
\end{equation}
The screened Coulomb interaction has a structure similar to
(\ref{barespinb}), since it, too, couples only to charge densities,
and does not depend on spin densities.  Consequently, the only
component of the response function $\Pi$ that affects the
transresistivity is the one coupling charge density to charge density.
For the fully polarized case, it is given by Eq.\ (\ref{Pieq}). We now
turn to discuss this component in the other extreme case, that of zero
spin polarization.

In principle, one may envision several Chern-Simons transformations
that might be applicable to the fully unpolarized $\nu=1/2$ state. At
$\nu=1/2$ there are two flux quanta per a hole. One possible
transformation attaches four flux quanta to each hole, and allows an
interaction only between same-spin composite fermions. This
transformation creates a correlated wave function of same-spin holes,
and reduces the probability of finding two same spin holes very close
to one another \cite{Read}. It does not, however, do the same for
holes of different spin directions. Thus, while the Coulomb energy
cost associated with holes of identical spins is reduced, that of
holes of different spin directions is unaffected.  An energetically
favorable transformation, that reduces the Coulomb energy of both
types of interactions, is one that attaches two flux quanta to each
hole, and allows an interaction that is independent of the spin
direction. The correlations induced by this transformation are, at
least on the RPA level, spin-independent. We use the second
transformation.

The calculation of $\Pi$ is similar to that leading to Eq.\ 
(\ref{pizz}). Eq.\ (\ref{picf}) still holds, although ${\tilde\Pi}^{e
  }, {\tilde C}$ and ${\tilde \Pi}^{\mbox {\tiny CF}}$ become single
layer $4\times 4$ matrices, with the entries corresponding to
densities and transverse current of two spin states. Within mRPA, in
the basis of eigenstates of $\sigma_z$, the response function
$\Pi^{\mbox {\tiny CF}}$ is a block-diagonal matrix with two pairs of
identical blocks, denoted by ${\tilde \chi}^{\mbox {\tiny CF}} \equiv
{\tilde \Pi}^{\mbox {\tiny CF}}_{\uparrow\uparrow} = \tilde\Pi^{\mbox
  {\tiny CF}}_{\downarrow\downarrow}\ $ (Note that
${\tilde\chi}^{\mbox {\tiny CF}},\tilde\Pi^{\mbox {\tiny
    CF}}_{\uparrow\uparrow}, \tilde\Pi^{\mbox {\tiny
    CF}}_{\uparrow\uparrow}$ are $2\times 2$ matrices with entries
corresponding to density and transverse current).  Since in our choice
of the Chern-Simons transformation the Chern-Simons interaction is
independent of the spin direction, it is again more convenient to
transform the spin states from eigenstates of $\sigma_z$ to
$\rho_\uparrow\pm\rho_\downarrow$ (and similarly for transverse
currents).  The components relevant for our calculation are the ones
coupling charge density and currents (i.e., spin density and currents
are irrelevant), and they are given by the $2\times 2$ matrix
$2\tilde\chi^{\mbox {\tiny CF}}$. Using Eq.\ (\ref{picfrpaa}) and
(\ref{picfrpab}), we then find the following expression for the
electronic density-density response function, to be compared with Eq.\ 
(\ref{pizz}),
\begin{equation}
  \Pi({\bf q},\omega) = \frac{2\chi^{\mbox {\tiny CF}}_{00}}{1 - 
    4\chi^{\mbox {\tiny CF}}_{00}\chi^{\mbox {\tiny CF}}_{11} 
    \left(\frac{2\pi\tilde\phi\hbar}{q}\right)^2} \, . 
  \label{Pieq_th}
\end{equation}

Qualitatively, Eq.\ (\ref{Pieq_th}) is very similar to Eq.\ 
(\ref{pizz}), the corresponding response function for the polarized
state. Quantitatively, there are three differences: a factor of $2$ in
the numerator resulting from the presence of two spin directions, a
factor of $4$ originating from the interaction of each composite
fermion with flux tubes carried by fermions of $two$ spin directions,
and a factor $1/\sqrt{2}$ difference between ${\rm Im}\chi_{\mbox
  {\tiny CF}}$ and ${\rm Im}\Pi_{\mbox {\tiny CF}}$, originating from
the difference between the Fermi wave-vector at the polarized and
unpolarized states.

We note, by passing, that this quantitative difference between $\Pi$
of the fully unpolarized state and $\Pi$ of the fully polarized state
should be observable in measurements of the $q$-dependent
conductivity using surface acoustic waves.  For example, one might
imagine a measurement of surface acoustic waves attenuation in a
$\nu=3/2$ state as a function of a magnetic field applied parallel to
the two dimensional electronic system. If the parallel magnetic field
modifies the spin polarization of the $\nu=3/2$ state, a significant
change in $\sigma_{xx}({\bf q})$ should be observable.

Coming back to the transresistivity, an examination of Eq.\ 
(\ref{Pieq_th}), Eq.\ (\ref{impiv}) and Eq.\ (\ref{donald}) reveals
that the drag transresistivity in the unpolarized case can be obtained
from that of the polarized case by three modifications: firstly,
$\tilde\phi$ should be multiplied by $2$, to account for the factor
$4$ in the denominator of (\ref{Pieq_th}). Secondly, the bare
interaction should be multiplied by $2$, to account for the factor $2$
in the numerator of (\ref{Pieq_th}) . And thirdly, the Fermi momentum
$k_{\scriptscriptstyle F}$ appearing in the electronic density-density
response function (\ref{pizz}) should be divided by $\sqrt{2}$.
Neglecting $\alpha$ compared to $1$, we then find that the
transresistivity in the unpolarized case is larger, by a factor of
$2^{2/3}$ compared to that of the polarized case.

\section{Summary}
\label{summary}

In this paper we considered the Coulomb drag between two layers of two
dimensional electron gases in a strong magnetic field, when the
filling fraction in each layer is at, or close, to $\nu=1/2$,
$\nu=1/4$ and $\nu=3/4$, and the electrons in each layer form a
compressible state. Using the composite fermion approach to analyze
the properties of each layer, and an electronic approach to analyze
the coupling between the layers, we find that the unique linear
$q$-dependence of the longitudinal conductivity, $\sigma(q)\propto q$,
at and close to even denominator filling fractions, lead to a unique
temperature dependence of the transresistivity between two layers,
$\rho_{\scriptscriptstyle D}\propto T^{4/3}$. In contrast to previous
works, we do not associate this temperature dependence to the nature
of the liquid formed by composite fermions, but rather interpret it as
a result of the slow relaxation of density fluctuations in a
compressible partially filled Landau level.

We examine the transresistivity at $\nu=3/2$ and find that it depends
significantly on the spin polarization of the $\nu=3/2$ state. We
therefore propose that measurements of $\rho_{\scriptscriptstyle D}$
at $\nu=3/2$ might shed light on that polarization, which so far is
not well understood.

Our results rely on the assumption of weak coupling between the layers
and neglect the finite thickness of each layer. The properties of each
layer are analyzed using approximation schemes for the composite
fermion problem. We find that $\rho_{\scriptscriptstyle D}$ depends
mostly on the strength of the Coulomb interaction and on geometrical
factors, such as the inter-layer and inter-electron distance. Its
dependence on masses, whether bare or effective, is rather weak. We
believe that these observations make the results we obtain mostly
independent of the fine details of the approximation we use.

\acknowledgements

We are indebted to J.~P. Eisenstein and T.~J. Gramila for discussing
unpublished data with us, and to B.~I. Halperin, A. Kamenev, Y. Oreg,
D. Orgad, and S.~H. Simon for instructive discussions. This research
was supported by a grant from the US-Israel Binational Science
Foundation and by a grant from the Minerva foundation (Munich,
Germany). A.~S. is supported by the V.  Ehrlich career development
chair.

\appendix

\section{The relation of composite fermion drag to electron drag}

In the main body of the paper we discussed the transresistivity from a
purely electronic point of view, showing that it originates from the
Coulomb drag between electrons at the two layers. In this appendix we
relate the electronic calculation to a calculation in terms of
composite fermions, which is the approach taken in Refs.\ \cite{KM96}
and \cite{S96}.

First, we note that as long as the inter-layer coupling is weak,
composite fermions in one layer do not interact with the Chern-Simons
field of the other layer (an inter-layer Chern-Simons interaction
gives rise to $331$-type states). Under these conditions, the
electronic drag resistivity is identical to the composite fermion drag
resistivity,
\begin{equation}
  \rho_{\scriptscriptstyle D}^e = 
  \rho_{\scriptscriptstyle D}^{\mbox {\tiny CF}} \, .
  \label{rer}
\end{equation}
In a single layer it is well known \cite{HLR93} that the composite
fermion resistivity matrix $\rho^{\mbox {\tiny CF}}$ and the
electronic resistivity matrix $\rho^{e}$ are related by
\begin{equation}
  \rho^e=\rho^{\mbox {\tiny CF}}+\frac{h}{e^2}\left (\begin{array}{cc}
      0 & \tilde\phi \\
      -\tilde\phi & 0
    \end{array} \right) \, .
\label{rescfre}
\end{equation}
Eq.\ (\ref{rer}) results from an extension of Eq.\ (\ref{rescfre}) to
the $4\times 4$ double layer resistivities. Similar to the single
layer case, electronic longitudinal resistivities are equal to those
of the composite fermions. The electronic (measurable) drag
conductivity is {\em not} identical to the composite fermion drag
conductivity.

Thus, in a composite fermion approach, we calculate
$\rho_{\scriptscriptstyle D}^{\mbox {\tiny CF}}$. The following
derivation is limited to the case of two identical layers at $\nu =
1/2$, and assumes that $\tilde \Pi^{\mbox {\tiny CF}}$ is diagonal.
Extensions will be discussed elsewhere \cite {US97}. Similarly to the
usual case, the transresistivity is given by \cite{ZH93,KM96,S96},
\end{multicols}
\renewcommand{\theequation}{A\arabic{equation}}
\widetext

\noindent
\begin{picture}(3.375,0)
  \put(0,0){\line(1,0){3.375}}
  \put(3.375,0){\line(0,1){0.08}}
\end{picture}
\begin{equation}
  \label{3:rho_D:cf}
  \rho_{\scriptscriptstyle D}^{\mbox {\tiny CF}} =
  \frac{1}{2(2\pi)^2} \frac{h}{e^2} \frac{1}{T  n^2}
  \int \frac {{\rm d} {\bf q}}{(2 \pi)^2} q^2
  \int_0^\infty \frac {\hbar{\rm d} \omega}{\sinh^2 \frac{\omega}{2 T}}
  \sum_{ij} {\rm Im} \Pi^{\mbox {\tiny CF}}_{ii} (q, \omega)
  {\rm Im} \Pi^{\mbox {\tiny CF}}_{jj} (q, \omega)
  \left| \left( U^{\mbox {\tiny CF}}_{\rm sc} \right)_{ij}
    (q, \omega) \right|^2,
   \label{rdcf}
\end{equation}
\hfill
\begin{picture}(3.375,0)
  \put(0,0){\line(1,0){3.375}}
  \put(0,0){\line(0,-1){0.08}}
\end{picture}

\begin{multicols}{2}
\noindent
where $i,j$ get the values $0$ or $1$, corresponding to density and
transverse currents. The summation over both density and transverse
current components is needed since the screened interaction couples
composite fermions both through their densities and their transverse
currents.  In Eq.\ (\ref{rdcf}) the composite fermion transresistivity
is given in terms of composite fermion response functions. We now turn 
to express it in electronic terms.
\end{multicols}

\pagebreak
\widetext

\begin{figure}
  \narrowtext
  \setlength{\unitlength}{0.6in}
  \begin{center}
    \begin{picture}(4.75,1)(0,-1)
      \put(0,-1){\psfig{figure=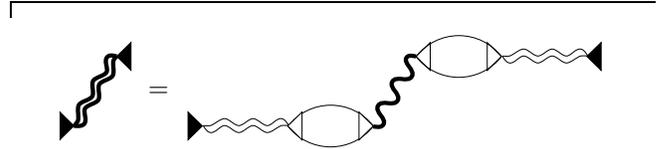,width=4.75\unitlength}}
      \put(0.875,-0.5){\makebox(0,0){$=$}}
    \end{picture}
  \end{center}
  \caption{
    The relation between $(U^{\mbox {\protect\tiny CF}}_{\rm
      sc})_{11}$ and $U^e_{\rm sc}$ [Eq.\ 
    (\protect{\ref{3:UUrelation}})] in diagrammatic terms. The bold
    double wavy line is the composite fermion screened interaction.
    The bold single wavy line is the screened Coulomb interaction.
    The thin double wavy lines are the bare Chern-Simons interaction,
    and the bubbles are electron response functions.  Density
    (transverse current) vertices are empty (full), and the two layers
    are represented by the two vertical positions in the diagrams. To
    obtain this relation diagrammatically, $(U^{\mbox {\protect\tiny
        CF}}_{\rm sc})_{11}$ and $U^e_{\rm sc}$ are represented as a
    sum over chains of bubbles. Each term in $U^e_{\rm sc}$ must start
    and end with a Coulomb interaction line. Each term in $(U^{\mbox
      {\protect\tiny CF}}_{\rm sc})_{11}$ must start and end with a
    transverse current vertex attached to a Chern-Simons interaction.
    The relation of the other elements of $U^{\mbox {\protect\tiny
        CF}}_{\rm sc}$ to $U^e_{\rm sc}$ may be represented in a
    similar fashion. }
  \label{plot2}
\end{figure}

We first relate the screened interlayer interaction of composite
fermions $U^{\mbox {\tiny CF}}_{\rm sc}$ to the electronic screened
interlayer interaction $U^e_{\rm sc}$ (the former is a $2 \times 2$
matrix, with density and transverse current indices, while the latter
is not, since electrons interact only through the Coulomb
interaction). By comparing the two, one finds
\begin{equation}
  \label{3:UUrelation}
  U^{\mbox {\tiny CF}}_{\rm sc} = \left(
    \begin{array}{ccc}
      (1 - C_{01} \Pi^e_{10}) U^e_{\rm sc}
      (1 - \Pi^e_{01} C_{10}) & & 
      - (1 - C_{01} \Pi^e_{10}) U^e_{\rm sc}
      \Pi^e_{00} C_{01} \\
      - C_{10} \Pi^e_{00} U^e_{\rm sc}
      (1 - \Pi^e_{01} C_{10}) & &
      C_{10} \Pi^e_{00} U^e_{\rm sc} \Pi^e_{00} C_{01}
    \end{array}
  \right) \, ,
\end{equation}
where the Chern-Simons interaction matrix $\tilde C$ is defined in
Eq.\ (\ref{cs}). This relation may be observed diagrammatically as in
Fig.\ \ref {plot2}, and may be derived from the expression for
$U^e_{\rm sc}$ [Eq.\ (\ref{interscg})],
\begin{equation}
  \label{A:Ue}
  U^e_{\rm sc} = 
  \frac {U_b}{(1 + \Pi^e_{00} V_b)^2 - (\Pi^e_{00} U_b)^2} \, ,
\end{equation}
where $V_b$ and $U_b$ are the intralayer and interlayer bare Coulomb
interactions. Inserting Eq.\ (\ref {A:Ue}) in the right hand side of
Eq.\ (\ref {3:UUrelation}), and using [see Eq.\ (\ref {picf})]
\begin{equation}
  \label{A:Picf}
  \tilde \Pi^e = \frac {1}{1 - \Pi^{\mbox {\tiny CF}}_{00}
    \Pi^{\mbox {\tiny CF}}_{11} | C_{01} |^2} \left(
    \begin{array}{cc}
      \Pi^{\mbox {\tiny CF}}_{00} &
      - \Pi^{\mbox {\tiny CF}}_{00}
      \Pi^{\mbox {\tiny CF}}_{11} C_{01} \\
      - \Pi^{\mbox {\tiny CF}}_{00}
      \Pi^{\mbox {\tiny CF}}_{11} C_{10} &
      \Pi^{\mbox {\tiny CF}}_{11}
    \end{array}
    \right) \, ,
\end{equation}
we obtain for the right hand side of Eq.\ (\ref{3:UUrelation}),
\begin{equation}
  \label{A:Ucf} 
  \frac {1}{(1 + \Pi^{\mbox {\tiny CF}}_{00} V_b - 
    \Pi^{\mbox {\tiny CF}}_{00}
    \Pi^{\mbox {\tiny CF}}_{11} | C_{01} |^2 )^2 -
    ( \Pi^{\mbox {\tiny CF}}_{00} U_b )^2 } \left(
  \begin{array}{cc}
    U_b &
    - U_b \Pi^{\mbox {\tiny CF}}_{00} C_{01} \\ 
    - C_{10} \Pi^{\mbox {\tiny CF}}_{00} U_b &
    C_{10} \Pi^{\mbox {\tiny CF}}_{00} U_b 
    \Pi^{\mbox {\tiny CF}}_{00} C_{01}
  \end{array}
  \right) \, .
\end{equation}
This is exactly the expression for $U^{\mbox {\tiny CF}}_{\rm sc}$
\cite {Bon}.

Using the relation (\ref {3:UUrelation}), the expression for the drag
resistivity becomes
\begin{equation}
  \label{3:rho_d:halfway}
  \rho_{\scriptscriptstyle D}^{\mbox {\tiny CF}} =
  \frac {1}{2 (2 \pi)^2} \frac {h}{e^2} \frac {2}{T n^2}
  \int \frac {{\rm d} {\bf q}}{(2 \pi)^2} q^2
  \int \frac {{\rm d} \omega}{\sinh^2 \frac{\omega}{2 T}}
  [\Xi (q, \omega)]^2
  | U^e_{\rm sc} (q, \omega) |^2 \, ,
\end{equation}
\hfill
\begin{picture}(3.375,0)
  \put(0,0){\line(1,0){3.375}}
  \put(0,0){\line(0,-1){0.08}}
\end{picture}

\begin{multicols}{2}
\noindent
where
\begin{equation}
  \label{3:Xi}
  \Xi = |1 - C_{01} \Pi^e_{10}|^2
  {\rm Im} \Pi^{\mbox {\tiny CF}}_{00} +
  |C_{10} \Pi^e_{00}|^2
  {\rm Im} \Pi^{\mbox {\tiny CF}}_{11} \, .
\end{equation}
We now show that $\Xi$ is, in fact, the imaginary part of the
electronic density-density response function, $\Pi^e_{00}$. From Eq.\ 
(\ref {A:Picf}) we have $\Pi^e_{00} = |\Pi^e_{00}|^2 / \Pi^{\mbox
  {\tiny CF} *}_{00} - |\Pi^e_{01}|^2 / \Pi^{\mbox {\tiny CF}}_{11}$.
Using ${\rm Im} z^{-1} = - |z|^{-2} {\rm Im} z$, we obtain
\begin{equation}
  \label{3:ImPi00}
  {\rm Im} \Pi^e_{00} =
  \frac {|\Pi^e_{00}|^2}{| \Pi^{\mbox {\tiny CF}}_{00}|^2}
  {\rm Im} \Pi^{\mbox {\tiny CF}}_{00} +
  \frac {| \Pi^e_{01} |^2}{| \Pi^{\mbox {\tiny CF}}_{11}|^2} 
  {\rm Im} \Pi^{\mbox {\tiny CF}}_{11} \, .
\end{equation}
On comparing (\ref {3:Xi}) with (\ref {3:ImPi00}), it is left to show
that
\begin{equation}
  \label{3:toshow}
  \frac {|\Pi^e_{00}|^2}{|\Pi^{\mbox {\tiny CF}}_{00}|^2} =
  |1 - C_{01} \Pi^e_{10}|^2 \, ,
  \; \; \;
  \frac {|\Pi^e_{01}|^2}{| \Pi^{\mbox {\tiny CF}}_{11}|^2} =
  |C_{10} \Pi^e_{00}|^2 \, .
\end{equation}
Both relations are verified using (\ref {A:Picf}).

The expression for drag resistivity is therefore written in terms of
the electronic properties, in exactly the way it is written for two
identical layers at $B=0$ [Eq.\ (\ref{donald})],

\end{multicols}
\widetext

\noindent
\begin{picture}(3.375,0)
  \put(0,0){\line(1,0){3.375}}
  \put(3.375,0){\line(0,1){0.08}}
\end{picture}
\begin{equation}
  \label{3:rho_D:e}
  \rho_{\scriptscriptstyle D}^{\mbox {\tiny CF}} =
  \rho_{\scriptscriptstyle D}^e = 
  \frac{1}{2(2\pi)^2}\frac{h}{e^2}\frac{1}{T n^2}
  \int \frac{d{\bf q}}{(2\pi)^2}\int_0^\infty  
  \frac{\hbar\, d\omega}{\sinh^{2}{\frac{\hbar\omega}{2T}}} \,
  q^2 \,  [{\rm Im} \Pi ({\bf q},\omega)]^2 \,
  |U_{\rm sc}({\bf q},\omega)|^2\, .
\end{equation}

\begin{multicols}{2}
\noindent
The only response function which enters this formula is the electron
density-density response function $\Pi = \Pi^e_{00}$, and this is the
only place where properties of the half-filled Landau level affect the
drag resistivity.

\end{multicols}

\end{document}